\documentclass[a4paper, 11pt, twocolumn]{article}
\usepackage[margin=2.5cm]{geometry}
\usepackage[utf8]{inputenc}
\usepackage[T1]{fontenc}
\usepackage[english]{babel}
\usepackage{graphicx}
\usepackage{amsmath}
\usepackage{amssymb}
\usepackage{caption}
\usepackage{subcaption}
\usepackage{hyperref}
\usepackage{float}
\usepackage{wrapfig}
\usepackage{array}
\usepackage{csquotes}
\usepackage{dsfont}
\usepackage{multirow}
\usepackage[switch]{lineno}

\usepackage[style = nature]{biblatex}
\graphicspath{ {./figures} }

\begin{document}

\title{Improving Predictions of Convective Storm Wind Gusts through Statistical Post-Processing of Neural Weather Models}
\author{Antoine \textsc{Leclerc}$^{\ast,1,2}$, Erwan \textsc{Koch}$^{1}$, Monika \textsc{Feldmann}$^{3}$, Daniele \textsc{Nerini}$^{4}$,  \\  Tom \textsc{Beucler}$^{1,5}$\\  
\ \\   
$^{1}$ Expertise Center for Climate Extremes (ECCE), University of Lausanne, Lausanne, Switzerland\\  
$^{2}$ \'Ecole polytechnique, Institut Polytechnique de Paris, Palaiseau, France \\   
$^{3}$ Institute of Geography - Oeschger Centre for Climate Change Research, \\  University of Bern, Bern, Switzerland \\  
$^{4}$ Federal Office of Meteorology and Climatology MeteoSwiss, Locarno, Switzerland \\   
$^{5}$ Faculty of Geosciences and Environment, University of Lausanne,\\ Lausanne, Switzerland \\
$^{\ast}$ Corresponding author. Email address: \href{mailto:antoine.leclerc@polytechnique.org}{antoine.leclerc@polytechnique.org}} 

\begin{titlepage}
    \maketitle
\end{titlepage}

\newpage


\begin{abstract}
Issuing timely severe weather warnings helps mitigate potentially disastrous consequences. Recent advancements in Neural Weather Models (NWMs) offer a computationally inexpensive and fast approach for forecasting atmospheric environments on a 0.25$^\circ$ global grid. For thunderstorms, these environments can be empirically post-processed to predict wind gust distributions at specific locations. With the Pangu-Weather NWM, we apply a hierarchy of statistical and deep learning post-processing methods to forecast hourly wind gusts up to three days ahead. To ensure statistical robustness, we constrain our probabilistic forecasts using generalised extreme-value distributions across five regions in Switzerland. Using a convolutional neural network to post-process the predicted atmospheric environment's spatial patterns yields the best results, outperforming direct forecasting approaches across lead times and wind gust speeds. Our results confirm the added value of NWMs for extreme wind forecasting, especially for designing more responsive early-warning systems.
\end{abstract}


\section{Introduction}

\begin{figure*}[h!]
    \centering
    \includegraphics[width=1\linewidth]{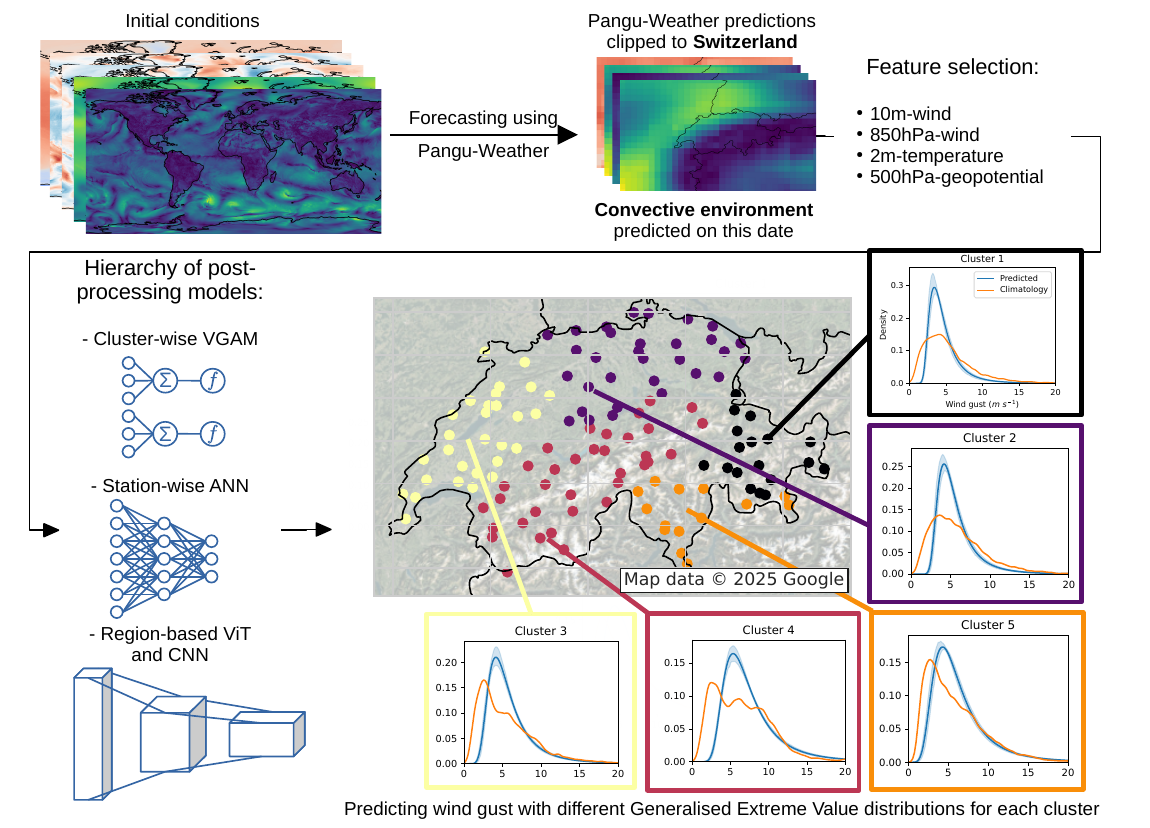}
    \caption{\centering Pangu-Weather forecasts for a day with convective conditions are clipped to Switzerland and post-processed using one of our different models. The best results are obtained with the CNN, which takes entire maps of features as input. Five different generalised extreme-value distributions are created, one for each cluster. This example shows predictions for June 2\textsuperscript{nd}, 2021 at 17:00 with a lead time of 27 h. Accuracy varies with the prediction (e.g., forecast for Cluster 2 is less accurate than for Cluster 4).}
    \label{fig:work-flow}
\end{figure*}

Severe winds can significantly damage buildings \cite{klawa2003model,pinto2007changing,donat2011high,prahl2012applying,prettenthaler2012risk} and pose risks to human lives \cite{schoen_climatology_2011,black_nontornadic_2010,ashley_spatial_2007}. Responding to such extreme events requires providing public authorities with accurate and easily understandable forecasts in time to implement prevention measures \cite{he2021wind}. Effective strategies rely on timely warnings, from long-term to short-term, and accurate medium-range forecasts (2--10 days) have proven particularly efficient when combined with impact forecasts \cite{schroeter_forecasting_2021}.

Meanwhile, machine learning-based models --- especially Neural Weather Models (NWMs) \cite{rasp2024weatherbench} --- are emerging as a viable complement to traditional numerical weather prediction (NWP) models \cite{ebert2023outlook} for short to medium-range weather forecasts. For example, GraphCast \cite{lam_graphcast_2023}, Pangu-Weather \cite{bi_accurate_2023}, FourCastNet \cite{pathak_fourcastnet_2022}, and AIFS \cite{lang2024aifs} can issue high-quality, deterministic and global forecasts with a 0.25$^{\circ}$ spatial resolution for medium range lead times (up to seven days). Once trained, these models can be run locally on a personal computer on a GPU in seconds. In contrast, physics-based NWP models solve non-linear partial differential equations using dedicated high-performance computing infrastructure. Although their outputs are generally made publicly available by national weather services and thus widely utilised, this complexity and cost prevent the general public from running them, hindering smaller weather prediction centres and hazard-issuing organisations.

The use of NWMs for severe weather forecasting is rapidly evolving, from tropical cyclone tracking \cite{bi_accurate_2023} to predicting local surface extremes \cite{price2025probabilistic}. However, forecasting thunderstorms — highly localised, short-lived, and inherently uncertain events — presents unique challenges. Although NWMs have typical spatial and temporal resolutions that are reasonable to forecast mesoscale environments conducive to severe weather \cite{feldmann2024convective}, they currently lack the resolution and variables (e.g., precipitation) needed to directly detect individual convective storms, which often last only a few hours and cover areas of just tens of square kilometres. To address this, we propose post-processing the atmospheric environments predicted by NWMs to forecast regional distributions of wind gusts (they can be seen as the distributions of possible wind gusts for a ``representative'' site within the region) rather than deriving the tracks of individual storms from forecasted fields. Although hail often remains the primary source of thunderstorm damage \cite{feldmann_hailstorms_2023}, we here focus on wind gusts as they depend on more predictable factors such as mean wind, shear, and topography \cite{sheridan_current_2018}, while still posing significant risks to infrastructure and public safety.

There are various methods to post-process NWP models to forecast station-wise wind gusts. For example, Friederichs et al. \cite{friederichs_forecast_2012} used Vector Generalised Additive Models (VGAMs) to predict station-wise Generalised Extreme-value (GEV) distributions of wind gusts and Vector Generalised Linear Models (VGLMs), Schulz et al. \cite{schulz_machine_2022} reviewed different statistical post-processing techniques (multi-linear regression, machine learning, and neural networks) to post-process wind gust distributions, and Fontana et al. \cite{fontana_convolutional_nodate} post-processed deterministic wind gusts forecasts via a Convolutional Neural Network (CNN) to predict distributions. As the considered wind gusts are maxima over 1 hour of the 1-s sustained wind speed (i.e., maxima over $3600$ observations), the GEV distribution, which is the asymptotic distribution of block maxima under appropriate conditions \cite[e.g.,][]{coles_introduction_2001}, appears as a natural model for them \cite[e.g., ][]{friederichs_forecast_2012,schulz_machine_2022}; their suitability in our context is expounded through diagnostic plots and Kolmogorov--Smirnov tests in the methods section. However, to our knowledge, these methods have not yet been applied to NWMs to predict regional distributions of meteorological variables.

Here, we post-process a NWM to issue probabilistic forecasts of regional hourly wind gusts using a hierarchy of empirical models (statistical and machine learning-based). We selected the Pangu-Weather NWM for this study because it is currently the only established NWM that generates medium-range forecasts with an hourly temporal resolution. Our study focuses on Switzerland, allowing us to leverage high-quality wind gust observations over complex topography to assess wind gust hazards under convectively active conditions. However, our framework can easily be adapted to other regions with station-based or gridded wind gust observations, enabling wind gust forecasts in areas where NWP models are not routinely run at the fine resolution needed to resolve extremes.

\section{Results}

\begin{figure*}[h!]
    \centering
    \includegraphics[width=1\linewidth]{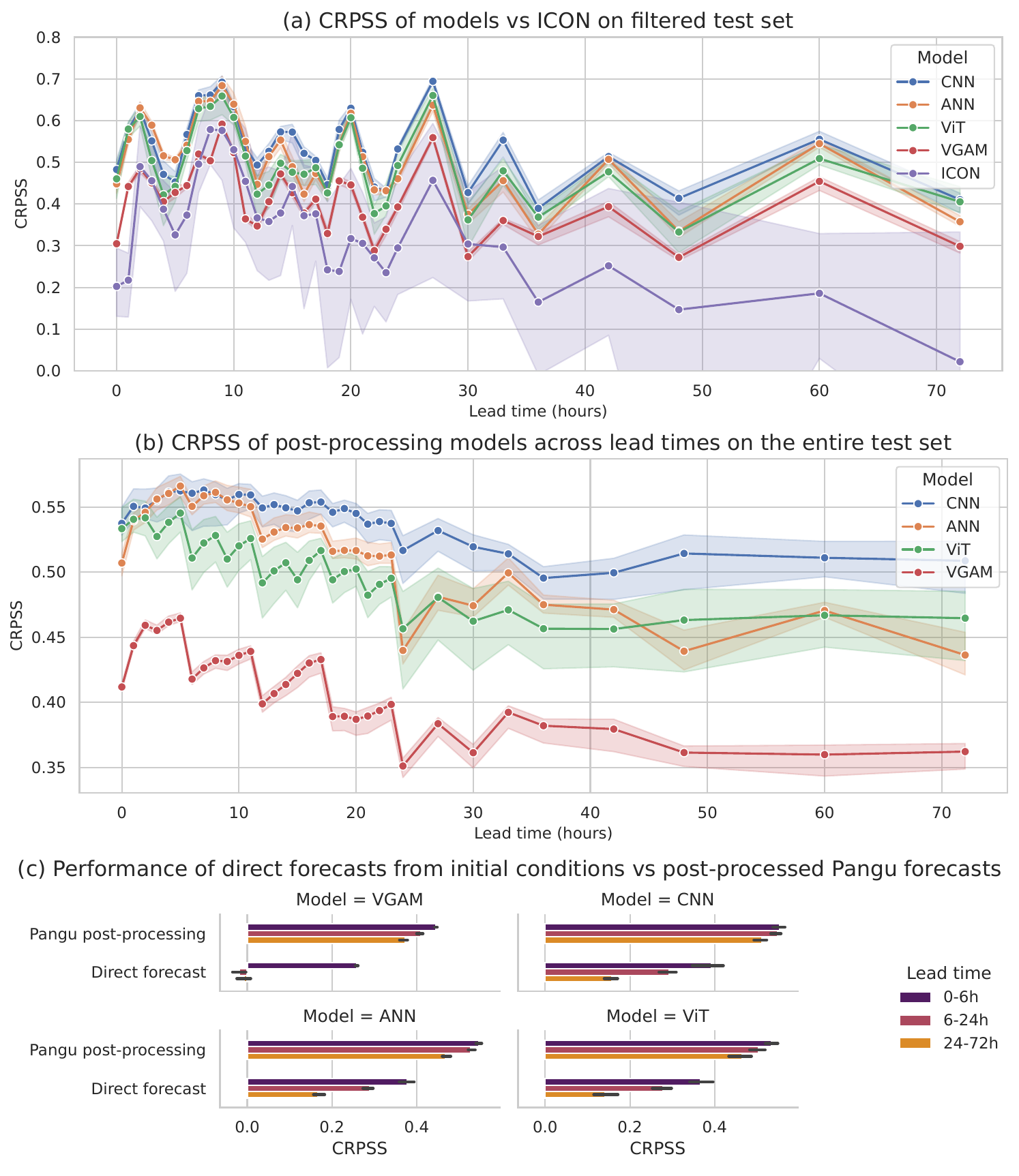}
    \caption{\centering For the specific task of predicting wind-gust GEV distributions over the five pre-defined Swiss meteorological regions, our post-processing models outperform direct forecasts, the ICON NWP reference, and the climatology of wind gusts across lead times, confirming the added value of Pangu-Weather (initialized with ERA5) in predicting atmospheric environments and the ability of statistically constrained, data-driven models to make reliable probabilistic wind gusts forecasts. Panels (a) and (b) show the mean CRPSS and its min/max over the five folds for different models as a function of lead time,  on a more restricted ensemble for (a) to use the same timestamps as ICON. Panel (c) shows the mean CRPSS and its min/max over the five folds for different models and different lead times, comparing a direct forecast to a post-processing of a Pangu-Weather forecast.} 
    \label{fig:general-score}
\end{figure*}

\subsection*{General performance}

Four post-processing models were trained on five years of Pangu-Weather forecasts (2016--2020) following a k-fold cross-validation scheme with each fold corresponding to one year in the training set. Each model predicts the location, scale, and shape parameters of five wind gust GEV distributions for each time step and each lead time, corresponding respectively to five clusters of stations from the Swiss Meteorological Network (SwissMetNet,  \cite{noauthor_automatic_2018}). The clusters are pre-computed based on measurement correlation between stations (see Section \ref{clustering} for more details). Performing a forward sequential feature selection with a VGLM led us to post-process only four fields forecasted by the NWM, namely 10 m-wind ($u_{10m}$), 2 m-temperature ($T_{2m}$), geopotential at 500 hPa ($\phi_{500hPa}$), and wind at 850 hPa ($u_{850hPa}$) (see Section~ \ref{seqfeatsel}). The models considered were then evaluated on the year 2021, filtered to keep only dates when Switzerland experienced a thunderstorm to account for convective activity. The models include a VGAM based on the \verb|evgam| \cite{youngman_evgam_2020} R package, an Artificial Neural Network (ANN) trained on the bilinear interpolation of Pangu-Weather forecasts on SwissMetNet stations, and a CNN and a Vision Transformer (ViT) trained on Pangu-Weather outputs clipped to Switzerland (Figure \ref{fig:work-flow}). All Neural Networks (NNs) were implemented in Python with the \verb|jax| library \cite{bradbury_jax_2018}.

Neural networks were trained using the Continuous Ranked Probability Score (CRPS) \cite{friederichs_forecast_2012} as a loss function, and the VGAM was calibrated by maximum likelihood estimation. To compare the different models across datasets, another metric was introduced: the  Continuous Ranked Probability Skill Score (CRPSS), defined as
\begin{multline}
\mathrm{CRPSS} \ =\\
\frac{\mathrm{CRPS} - \mathrm{CRPS}_{\mathrm{climatology}}}{\mathrm{CRPS}_{\mathrm{best \ possible}} - \mathrm{CRPS}_{\mathrm{climatology}}},
\end{multline}
where ``climatology'' corresponds to the empirical distribution of wind gusts across stations and times for each cluster, obtained with all wind gust measurements from April to October for years 2016 to 2020, and ``best possible'' refers to the empirical distribution of wind gusts on each cluster at a specific time. As the models aim at fitting a distribution in a cluster of stations, the ``best possible'' CRPS cannot be outperformed. Therefore, CRPSS varies from negative values to 1 (best possible model), with 0 corresponding to climatology (no skill). To benchmark post-processing models against a state-of-the-art NWP system, we also evaluated the direct output of the ICON-CH2-EPS ensemble (referred to as ICON). For this, MeteoSwiss provided operational 10-m wind gust forecasts for the test year 2021 (see Supplementary Figure 8, Supplementary Information section 5 for more details about how the test year was chosen).

Encouragingly, Figure \ref{fig:general-score}a shows that not only all post-processing models are skilful, but also outperform the ICON model, particularly at longer lead times. The gap between our best post-processing model (the CNN) and ICON is 0.2 for lead times under 32 hours but exceeds 0.3 for longer lead times. Pangu-Weather was initialized with ERA5 reanalysis, but previous research has shown that changing from ERA5 reanalysis to IFS operational analysis has little impact on the outcome of Pangu-Weather forecasts \cite{feldmann2024convective}, in line with NWMs' ensembles being underdispersive compared to traditional ensemble methods \cite{bi_accurate_2023}. We thus expect our results to be robust with regard to the choice of initial conditions. The similar sawtooth pattern for all models results from the use of different test sets at each lead time to accommodate the requirement of having a thunderstorm on the forecast date and the constraint of ICON being launched every 18 hours (see Section~\ref{sec_icon}).

Figure \ref{fig:general-score}b shows that all models substantially outperform the cluster-wise climatology with CRPSS always exceeding 0.35. The neural networks outperform the VGAM, with a mean relative improvement of 27\%. The best model in terms of CRPSS is the CNN, although all NNs perform similarly for short lead times. The performance of all models decreases with lead time: e.g., the CRPSS values decline by 12$\%$ on average between lead times less than 6 hours and larger than 24 hours. Finally, the sawtooth patterns are inherited from Pangu-Weather's approach to handling different lead times (see Supplementary Figure 1, SI section S1 for more details about the oscillations, and Supplementary Figure 2, SI section S2 for an investigation on CRPSS apparent stabilization at longer lead times).

\subsection*{Direct forecast vs post-processing Pangu-Weather}

To assess whether post-processing Pangu-Weather provides information for wind gust prediction or not, as NNs could be powerful enough to issue gust forecast from raw observations, other models with the same architecture were trained as simple forecasting baselines from the prediction dates with ERA5 dataset as initial conditions. Regional distributions for time $t + \Delta t$ were forecasted from ERA5 data at time $t$. Panel~(c) depicts the variation of CRPSS across lead times for all four ERA5-based models against the Pangu-based ones --- for which the forecasts for time $t + \Delta t$ were obtained by post-processing a Pangu-Weather forecast for lead time $\Delta t$ computed from ERA5 data at time $t$.

All models perform better when based on Pangu-Weather forecasts than on ERA5 initial conditions, confirming the added information of NWMs for our wind gust prediction cases. This is especially true for the ``direct forecast'' VGAM which behaves like the climatology at lead times  larger than 6 h. For the neural networks, the improvement also increases with lead time (mean relative CRPSS improvement of 44\% for the first 6 h, 84\% between 6 h and 24 h, and 215\% between 24 h and 72 h).

\subsection*{Assessing performance across wind gust speeds}

\begin{figure}[h!]
    \centering
    \includegraphics[width=1\linewidth]{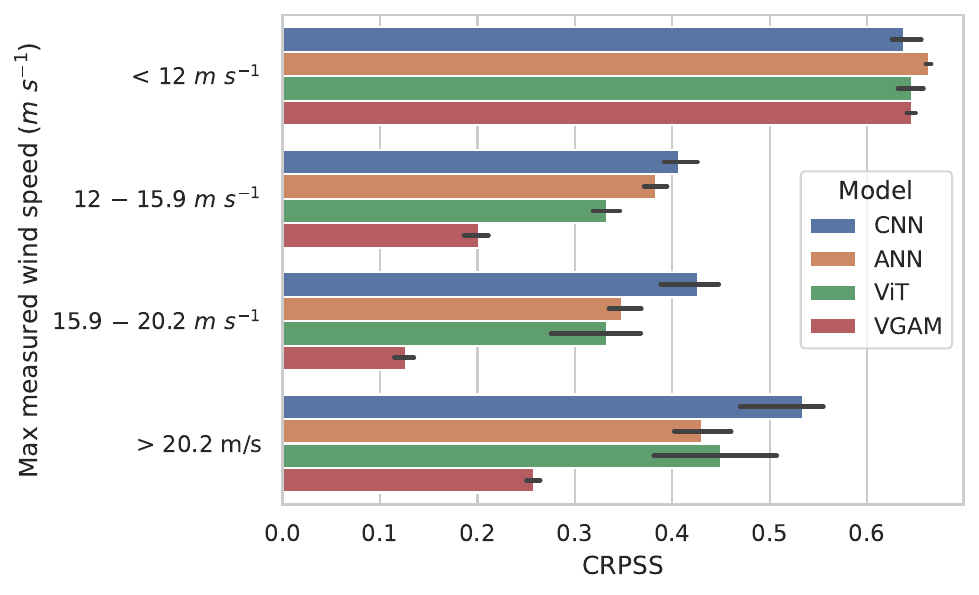}
    \caption{The CNN captures extremes better than the other models. The panel shows, for different categories of observed maximum wind gusts over the targeted region, the averaged CRPSS on all lead times and its min/max over the five folds, for all four models.}
    \label{fig:quantiles}
\end{figure}

The evaluation dataset (year 2021) was divided into four distinct subsets, each defined by the quantiles at levels 0.5, 0.75, and 0.9 of the maximum wind gust within each cluster: less than 12 m s$^{-1}$, between 12 m s$^{-1}$ and 15.9 m s$^{-1}$, between 15.9 m s$^{-1}$ and 20.2 m s$^{-1}$, and greater than 20.2 m s$^{-1}$. These wind gusts will be referred to as light, moderate, strong, and extreme, respectively (see Table \ref{tab:thresholds}).

\begin{table}[]
    \centering
    \begin{tabular}{c|c|c}
        Category & Wind gust (m s$^{-1}$) & Quantile \\ \hline
       Light  & \textless 12 & \textless 0.5 \\
Moderate  & 12--15.9 & 0.5--0.75 \\
       Strong & 15.9--20.2 & 0.75--0.9 \\
       Extreme & \textgreater 20.2 & \textgreater 0.9\\
    \end{tabular}
    \caption{Wind gust categories.}
    \label{tab:thresholds}
\end{table}


Figure \ref{fig:quantiles} shows that, except with light wind gusts, the CNN is the best model. The VGAM experiences a sharper decrease in its performance for higher wind gusts than other models (69\% decrease on average for the VGAM vs. 29\% for the CNN, 42\% for the ANN, and 42\% for the ViT on moderate, strong, and extreme winds against light wind). Interestingly, all models perform better for extreme wind gusts than for moderate and strong ones (see Supplementary Figures 3 to 5, SI section S3 for investigation on sensitivity to filtering). Extreme gusts are driven by specific conditions with high instability, and strong mid-level winds leading to fast-moving storms. These may be clearer to identify than intermediate cases producing less intense gusts (see Figure \ref{fig:shap}c).

\subsection*{Exploitation of the spatial context by the CNN}

\begin{figure*}[h!]
    \includegraphics[width=1\linewidth]{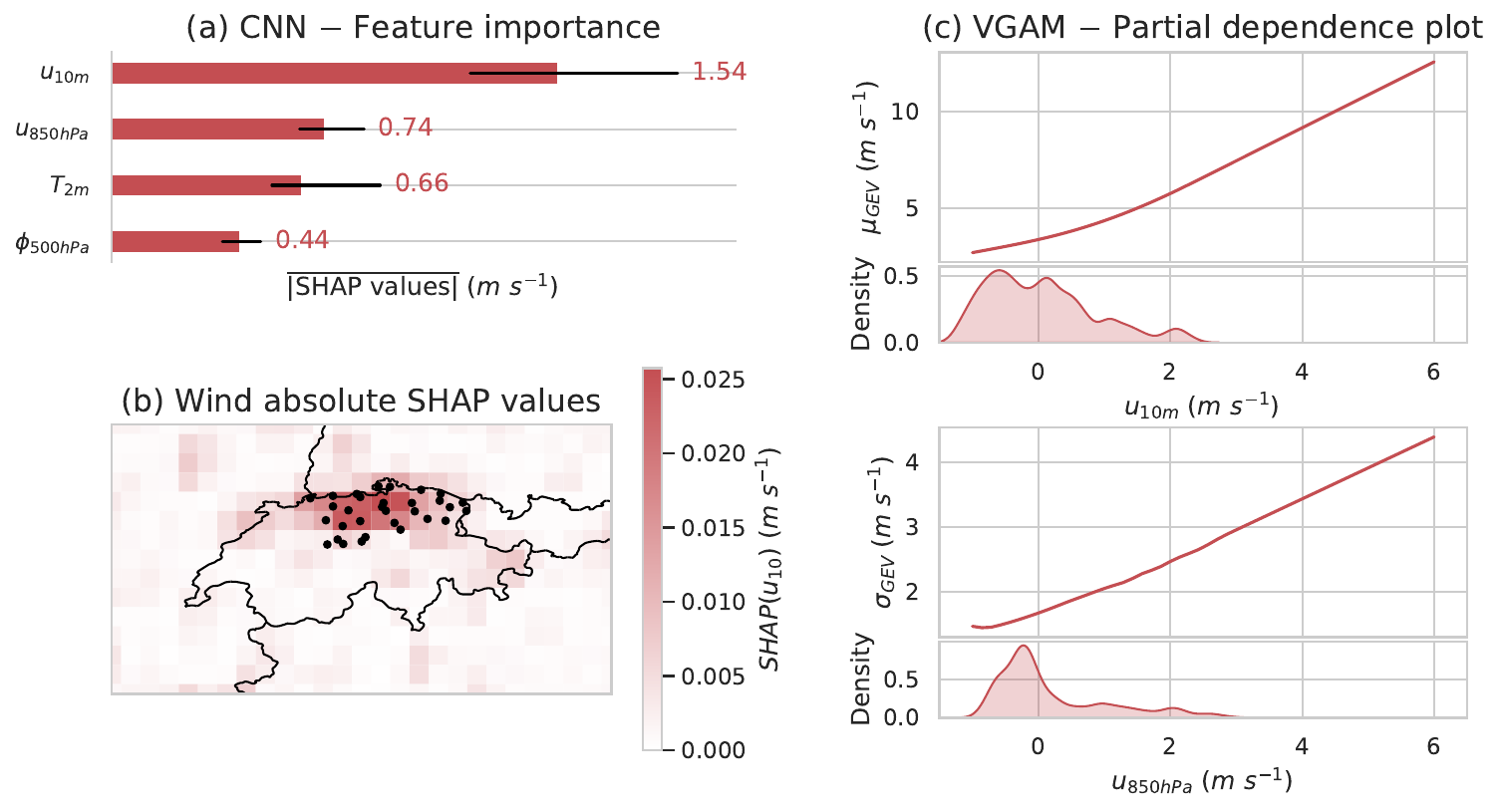}
    \caption{\centering Near-surface wind speed, near-surface temperature, 850 hPa wind and 500 hPa geopotential are used across models to predict wind gust distributions over each cluster --- a task for which the CNN primarily focuses on local information. Panel (a) shows the mean absolute SHAP value and its min/max for each feature of the CNN over the five folds, based on a random sample of 50 dates for cluster 1 and for the prediction of the location parameter $\mu$ of wind gusts. Panel (b) shows the mean absolute SHAP value for $u_{10m}$, based on the same samples (black dots represent the stations of cluster 1). Panel (c) represents the VGAM partial dependence plot for the two main predictors for $\mu$ and $\sigma$, along with the univariate distributions of said predictors. It shows how an increase in 10m wind implies more important wind gust, and how an increase in 850hPa winds implies more dispersion of wind gust distribution.}
    \label{fig:shap}
\end{figure*}

To gain a deeper understanding of our best post-processing models, we compute SHapley Additive exPlanations (SHAP, \cite{lundberg2017unified}) using the game theory-based \verb|shap| \cite{noauthor_welcome_nodate} interpretability library in Python. Shapley values quantify how each input feature shifts a specific model prediction away from its expected (mean) value. By comparing the mean absolute SHAP value associated with each feature, we can assess their relative importance, including that of meteorological variables and the spatial importance of pixels within their fields.

Figure \ref{fig:shap}a shows that the near-surface wind $u_{10m}$ remains the most important variable in terms of SHAP values to predict the location parameter $\mu$ of wind gusts, agreeing with the VGLM  feature selection results (see Section \ref{seqfeatsel} for more details). The $u_{10m}$ variable is also the one representing the most direct prediction of wind gusts. The $u_{850hPa}$ variable ranks higher in the CNN's feature importance than when sequentially selecting features of simpler models (VGAM), and shares similar importance with  $T_{2m}$. Lower tropospheric wind can lead to surface gusts via downward momentum transfer and is hence meteorologically related to wind gusts. The temperature $T_{2m}$ plays an important role in instability and thunderstorm environments overall. Finally, $\phi_{500hPa}$ is the least important feature to predict $\mu$. It here represents the large scale, indicating the presence of synoptic features, like an approaching front, that are related to potential convective hazards.

Figure \ref{fig:shap}b shows that the CNN exploits mainly local information inside the regional cluster to predict wind gusts (SHAP values larger than 0.20 m s$^{-1}$) but not only: more diffused information is used from other parts of the map (SHAP value larger than 0.05 m s$^{-1}$). 

Figure \ref{fig:shap}c provides insight into the contribution of $u_{10m}$ and $u_{850hPa}$ to wind gust forecasts. 
The wind gusts' location parameter increases with $u_{10m}$, with a coefficient of 1.4. This is significantly smaller than the theoretical wind gust factor for gusts measured over one second and mean wind measured over one hour, which has a theoretical value of 3.4 \cite{noauthor_gust_nodate}. The parameter $\sigma$ also increases with $u_{850hPa}$, with a coefficient of 0.5.

\section{Discussion}

Our results suggest that post-processing Pangu-Weather improves the prediction of convective storm wind gusts on short to medium-range lead times. The CNN provides the best results and outperforms all other models at lead times longer than 18 hours, including the NWP ICON. When filtering wind gusts to keep only cases with extreme convective winds, performance degrades for all models. However, all models still outperform climatology across lead times (0--72 hours), confirming the predictability of even extreme cases. Also, directly forecasting from ERA5 initial conditions substantially decreases the performance of all models, which suggests that Pangu-Weather contains wind gust-related information. Finally, the increasing performance gap with lead time between neural networks and the VGAM indicates that spatial patterns play a role in refining forecasts, especially at longer lead times when the spatial resolution of NWMs' forecasts decreases.

Feature importance analysis provides insight into how the models predict wind gusts. Since wind gusts are defined as extrema of the 10 m wind field, it is unsurprising that $u_{10m}$ is the most important feature in both the VGLM feature selection and the CNN SHAP value plot, with the location parameter for wind gusts increasing with $u_{10m}$. While the CNN primarily relies on wind values within the cluster, it interestingly uses larger spatial context at longer lead times. This aligns with the diminishing predictability of finer scales over time, while large-scale patterns remain predictable for longer. VGAM partial dependence plots show that the wind gust scale parameter increases with $u_{850hPa}$, reflecting greater variability in surface wind gusts due to stronger wind shear at higher altitudes and the uncertainty of whether high winds aloft will reach the surface.

Comparing the theoretical and fitted gust factors reveals a substantial discrepancy, with the theoretical value more than twice as large as that estimated by VGAM methods. While the VGAM incorporates additional predictors such as temperature, geopotential, and higher-altitude mean wind in an additive framework, some deviation from the theoretical gust factor is expected. However, the magnitude of this discrepancy confirms the insufficiency of mean wind alone and the importance of additional predictors for accurately estimating wind gusts.

In summary, post-processing methods applied to Pangu-Weather outputs support the idea that NWMs provide useful information for the prediction of local-scale extremes. In the case of convective wind storms, two successive filtering (dates with a storm detected in Switzerland and dates with wind gusts exceeding a determined threshold on a determined area) still provided significantly better results than a climatology-based estimation of wind gusts: not only are wind gusts a predictable proxy variable for convective storms, but forecasting their speed can be achieved through the post-processing of global AI models.

The inherent limitations of NWMs and convective storm forecasts can be addressed with distribution-based methods, which can handle uncertainties from convective storm dynamics and the coarse resolution of NWMs while maintaining regional predictability. Future work could explore temporal aggregation to mitigate timing uncertainties in convective events, such as deriving daily wind gust distributions by aggregating hourly forecasts. Testing alternative NN architectures could further improve performance, particularly for strong and extreme wind gusts, by incorporating loss functions that penalize errors on extreme gusts more heavily than those on lighter gusts.

The approach presented here is expected to generalise to other regions as well. The context of complex topography complicates thunderstorm and wind gust predictions in Switzerland, implying a possibly even better performance in less mountainous areas.

The practical application of Pangu-Weather post-processing could involve issuing severe wind gust warnings for specific regions or clusters. Since the models provide full distributions for each cluster, warnings could be based on a threshold applied to the predicted wind gust distributions. A current limitation of our approach is that we filtered the dataset based on the presence of a convective storm in Switzerland to focus on convective storm wind gusts. Future work could address this by developing real-time filtering methods based on convective parameters such as Convective Available Potential Energy (CAPE) and Convective Inhibition  (CIN) over target regions, which could be evaluated on data not used for training (e.g., 2022 and beyond) to further enhance the operational relevance of the proposed framework.

Finally, worthwhile future work would involve obtaining spatially-resolved forecasts based on our current approach. One option would be to draw a value from our forecast distribution and simulate at all sites within the cluster, conditioned on this value, using appropriate random fields or copula models.

\section{Methods}
\subsection*{Pangu-Weather model}
We chose to post-process Pangu-Weather to take full advantage of the hourly time resolution of its outputs. Pangu-Weather is an NWM consisting of a 3D Earth-specific transformer trained on the ERA5 dataset, a 43-year meteorological reanalysis provided by the European European Centre for Medium-Range Weather Forecasts \cite{bi_accurate_2023}. It provides forecasts of 4 surface variables and 5 upper air variables for 13 pressure levels (Table \ref{tab:PWvars}) at a spatial resolution of 0.25° $\times$ 0.25°. Pangu is a succession of four models which were each trained for a specific lead time (1h, 3h 6h, and 24h). To produce a forecast, a minimal number of Pangu submodels is used, from the longest to the shortest lead time. 

\begin{table}[h]
\centering
    \begin{tabular}{|c|p{3cm}|c|}
    \hline
    Type & Variable & Unit \\\hline
    \multirow{4}{1.2cm}{Surface} & 10m $u$-component of wind & m s$^{-1}$ \\
    & 10m $v$-component of wind & m s$^{-1}$ \\
    & 2m temperature & K \\
    & Mean sea-level pressure & Pa \\\hline
    \multirow{5}{1.2cm}{Upper air} & Geopotential & m s$^{-2}$ \\
    & Specific humidity& kg kg$^{-1}$\\
    & Temperature& K \\
    & $u$-component of wind & m s$^{-1}$ \\
    & $v$-component of wind & m s$^{-1}$ \\\hline
    \end{tabular}
    \caption{\centering Pangu-Weather variables on which feature selection is performed.}
    \label{tab:PWvars}
\end{table}

We used Pangu-Weather to produce a training set of 5 years (2016 to 2020) and a test set of one year (2021, on which Pangu-Weather was not trained) clipped to Switzerland, each year containing the months April to October as they correspond to the convective season described in \cite{feldmann_hailstorms_2023}. To balance resolution and cost, for each date, we produced forecasts with hourly lead times up to 1 day, 3-hourly lead times up to 2 days, and 12-hourly lead times up to 3 days. 

In addition to Pangu-Weather's standard variables (Table \ref{tab:PWvars}), two storm-related variables were calculated with the Python library \verb|wrf-python| \cite{technologies_geoscience_2024}: the maximum CAPE in the lowest layers to assess the environmental convective instability following the method from \cite{allen_severe_2011}, and the 0--3 km above ground storm-relative helicity (SRH) to assess tornadic instability and supercell presence \cite{kerr_storm-relative_1996}.

\subsection*{SwissMetNet observations}

Wind gusts were obtained through the SwissMetNet dataset \cite{noauthor_automatic_2018}, a network of 157 automatic stations disseminated over Switzerland providing measurements of different meteorological variables such as precipitation, temperature, air humidity, and wind every 10 minutes. Hourly measurements of wind gusts (in m s$^{-1}$) --- defined as the \textit{maximum wind speed over the preceding hour sustained during 1 second} --- were used as the target variable. The number of working SwissMetNet stations varies from one month to the other, as some  stations were added or repaired between 2016 and 2021. We alleviated this issue by filtering the stations so that only those working through at least 70\% of the dates in the dataset were kept. Finally, only the dates with complete station wind gust measurements data were kept.

In the end, the complete dataset contains 128 stations, 20,575 time steps, and 33 lead times. As the percentage of dates in the training set where a storm is detected by a station was a hyperparameter of the post-processing model, not all time steps were used for each model.

\subsection*{Storm detection}

The storm tracks provided in the dataset used by \cite{feldmann_hailstorms_2023} were used to evaluate the different models on severe events and to filter storm events in the training set. A supplementary filter on the storms was then added to keep only the ones detected by a station, as the dataset also contained storms detected by Swiss radar measurements in neighbouring countries.

\begin{figure}[h]
    \centering
    \includegraphics[width=\linewidth]{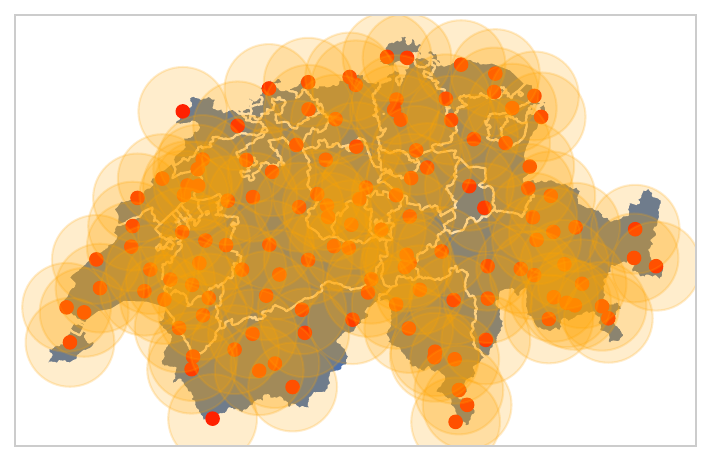}
    \caption{\centering Filtered stations (red dots) with 25 km radius (orange area) cover the whole Swiss territory (blue surface).}
    \label{fig:25km-radius}
\end{figure}

The criterion for a storm to be detected by a SwissMetNet station was defined as ``part of its trajectory stands within a 25 km radius of the station'' --- so that the whole Swiss territory is covered by the resulting area (Figure \ref{fig:25km-radius}).

A comparison of the thunderstorm dataset with SwissMetNet station coordinates also allowed us to find stations within a specific radius of each storm's track, which allowed us to perform a storm-specific evaluation of the performance of a wind gust forecast.

\subsection*{ICON dynamical baseline}
\label{sec_icon}

The ICON-CH2-EPS is a high-resolution ensemble prediction system utilized by MeteoSwiss for regional numerical weather forecasting in the Alpine region. It is based on the ICOsahedral Non-hydrostatic \cite[ICON][]{zangl2015icon} modelling framework, a scalable and flexible model designed for both operational and research applications in meteorology and climate science. In its operational configuration, the ICON-CH2-EPS runs on a 2.2 km horizontal grid spacing and executes 21 ensemble members four times daily, generating forecasts extending up to five days ahead.

As ICON-CH2-EPS did not run yet in 2021, the ICON data used in this study are part of a set of reforecasts covering April to October 2021 with a reduced update frequency (every $18$ h) and lead times ranging from 0 to 72 h. From the 21-member ensemble forecast, station-wise 10 m wind gust values - computed as hourly maximum of convective and turbulent components of wind gust - were obtained by finding the nearest pixel to the SwissMetNet stations on the 2.2 km grid.

To provide a fair comparison with our post-processing models, the test set was filtered to retain only ICON initialization dates for which a storm was detected over Switzerland on the corresponding forecast date.

\subsection*{Probabilistic scoring}

The Continuous Ranked Probability Score (CRPS) was chosen as a metric to train and evaluate the different models. As a proper scoring rule validating all three of Murphy's forecast goodness types \cite{murphy_what_1993}, it is widely used to assert weather model performance (\cite{bi_accurate_2023}, \cite{schulz_machine_2022}, \cite{bremnes_evaluation_2024}). Moreover, CRPS-based training has already provided the best forecasts for high-speed wind gusts in other studies \cite{friederichs_forecast_2012}.

The CRPS measures the squared difference between the forecast distribution function $F$ and the observation $y$:

\begin{equation}
    \mathrm{CRPS}\left(F,y\right) \ = \ \int_\mathbb{R} \left(F(x) - \mathbb{I}_{\{ y\leq x\}}\right)^2 \mathrm{d}x.
\end{equation}

where $\mathbb{I}_{\{ y\leq x\}}$ is the indicator function of $[y, +\infty[$.

It can be generalised for an empirical distribution of observations, and it can be shown that optimising $F$ for every observation considered individually is equivalent to optimising it compared to the empirical distribution.

\subsection*{SHAP values}

For each model $m$, the mean absolute SHAP value of a variable $u$ is defined by :
\begin{equation}
\begin{split}
|\overline{\text{SHAP}(u_m)}| \ = \\
 \sum_{(x,y,\Delta t, T) \in S} &| \text{SHAP} \big ( \ u_m(x,y,\Delta t, t) \ \big )  |
\end{split}
\end{equation}

where $u$ is the predicted variable, $x$ is the longitude, $y$ is the latitude, $\Delta t$ is the lead time, $T$ is the targeted prediction date, and $S$ is a sample of prediction date with all lead times, all longitude, all latitude.

Figure \ref{fig:shap} (a) shows the min-max of $|\overline{\text{SHAP}(u_m)}|$ for $m \in [1,2,3,4,5]$ being the five CNN parametrisation from the k-fold training.

For Figure \ref{fig:shap} (b), at each point, the SHAP value is given by :

\begin{equation}
\begin{split}
|\overline{\text{SHAP} \big(u(x,y\big)}| \ = \\
 \sum_{(\Delta t, T,m) \in R} &| \text{SHAP} \big ( \ u_m(\Delta t, t) \ \big )  |
\end{split}
\end{equation}

where $u$ is the predicted variable, $x$ is the longitude, $y$ is the latitude, $\Delta t$ is the lead time, $T$ is the targeted prediction date, and $R$ is a sample of prediction date with all lead times, all models.

\subsection*{Spectral clustering of SwissMetNet stations}\label{clustering}

Pangu-Weather, as most other AI-based NWMs, was trained with a coarse, 0.25° resolution. Studying small-scale phenomena from raw forecasts is thus rather challenging. It is more efficient and accurate to spatially aggregate measurements to compare them with larger scale forecasts.

To do so, the stations were grouped into different clusters. Using the absolute product of precipitation correlation and wind gust correlation between stations as a hybrid affinity measure, spectral clustering was applied to the 128 stations.

\begin{figure}[h]
    \centering
    \includegraphics[width=1\linewidth]{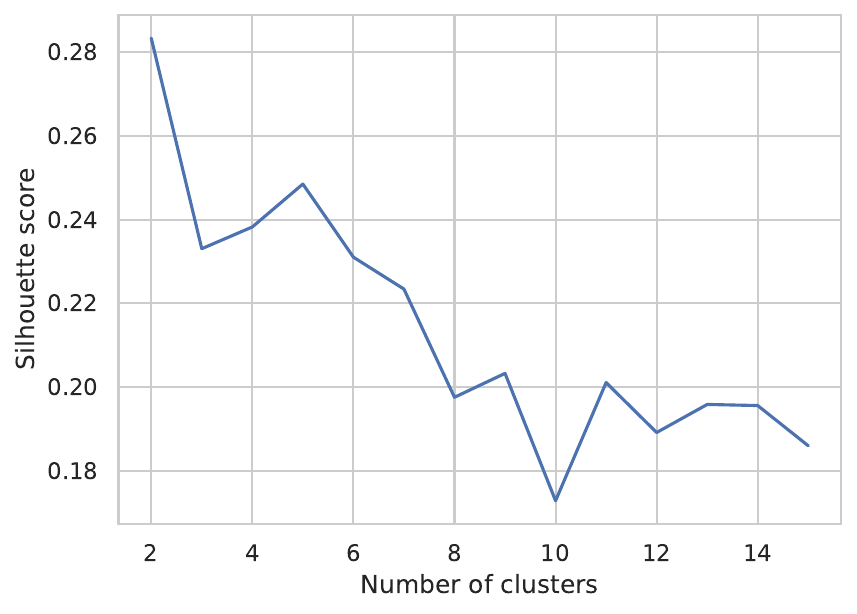}
    \caption{\centering 
    The silhouette score of spectral clustering exhibits a peak at $n_c = 5$. The plot represents the silhouette score as a function of the number of clusters.
    }
    \label{fig:silscore}
\end{figure}

Figure \ref{fig:silscore} shows that the silhouette score reaches its maximum for $n_c = 2$, where $n_c$ denotes the number of clusters, which corresponds to a division of the stations along a North-South axis. However, it also reaches a local maximum for $n_c = 5$. As local forecasts are more insightful for regional prevention of severe events, we chose to consider five smaller prediction regions rather than two even though we expected the results to be slightly less accurate.

\begin{figure}[h]
    \centering
    \includegraphics[width=1\linewidth]{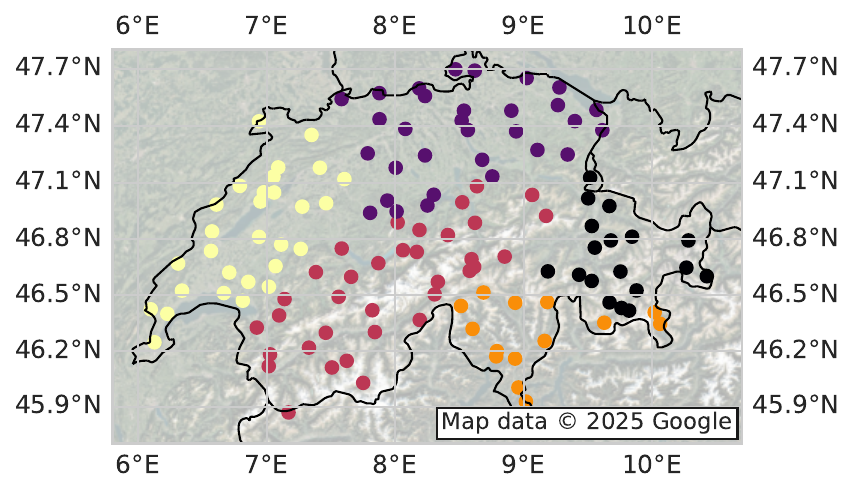}
    \caption{\centering
    Clusters for $n_c = 5$ are spatially and climatically coherent. The clusters have been obtained by the spectral clustering algorithm. The product of absolute wind gust and precipitation correlations between stations was used as an affinity measure.}
    \label{fig:cluster5}
\end{figure}

Interestingly, even though no direct spatial indicator was given as an input for the clustering (i.e., no indication of position), the obtained clusters are spatially coherent: they do not intersect, present climatic similarities, and approximately align with the regions used by MeteoSwiss to define the thresholds for extreme wind hazard warnings \cite{MeteoSwiss2021}. Figure \ref{fig:cluster5} reveals a clear separation between various homogeneous climate zones: the Alps (red for western Alps and black for eastern Alps), southern plains (orange), western plains and Jura (yellow), and northern plains (purple). The climatic stationarity observed in the obtained zones supports our assumption of constant values for $\mu$, $\sigma$, and $\xi$ across each region.

\subsection*{Generalised Extreme-Value distribution}

Extreme-value theory provides asymptotic statistical models for block maxima and threshold exceedances, enabling the estimation of probabilities for events more extreme than those observed in historical records. In the univariate case, the appropriate distribution for block maxima is the Generalised Extreme Value (GEV) distribution, characterized by a location parameter $\mu \in \mathbb{R}$, a scale parameter $\sigma > 0$, and a shape parameter $\xi \in \mathbb{R}$. Its cumulative distribution function $F_{\mu, \sigma, \xi} $ is given by

\begin{equation}
    F_{\mu, \sigma, \xi}(x) = 
    \exp\left\{ -\left[1 + \xi \left( \frac{x - \mu}{\sigma} \right) \right]^{-1/\xi} \right\},
    \label{Eq_GEVdf}
\end{equation}
for values of $x$ such that $1 + \xi (x - \mu)/\sigma > 0$. When $\xi = 0$, the distribution corresponds to the limit of \eqref{Eq_GEVdf} as $\xi \to 0$.

As wind gusts are defined as the maximum wind speed over a certain time period, modelling them using GEV distributions is often suitable (e.g, \cite{pinheiro_comparative_2016}, \cite{cheng_generalized_2002}, \cite{friederichs_probabilistic_2009}). Moreover, a notable advantage of GEV distributions in training machine learning models is the availability of a closed-form expression for the CRPS \cite{friederichs_forecast_2012} that can be implemented directly as a loss function:

For $\xi \neq 0$:

\begin{equation}
    \begin{split}
        C&RPS(F_{(\mu,\sigma,\xi)}, y) = 
        (\mu - y-\frac{\sigma}{\xi})(1-2F_{\mu,\sigma,\xi})\\
        &-\frac{\sigma}{\xi} (2^\xi \Gamma(1-\xi)- 2\Gamma_l(1-\xi, -\log F_{\xi,\sigma,\mu}(y)))
    \end{split}
\end{equation}

For $\xi = 0$:

\begin{equation}
    \begin{split}
        CRPS(F_{(\mu,\sigma,0)}, y) & = 
        \mu - y + \sigma(\gamma - \log 2)\\
        &- 2\sigma Ei(\log F_{\mu, \sigma, 0} (y))
    \end{split}
\end{equation}

\subsection*{Suitability of the GEV distribution}
\label{app_suit_gev}

\begin{figure*}[h!]
    \centering
    \includegraphics[width=\linewidth]{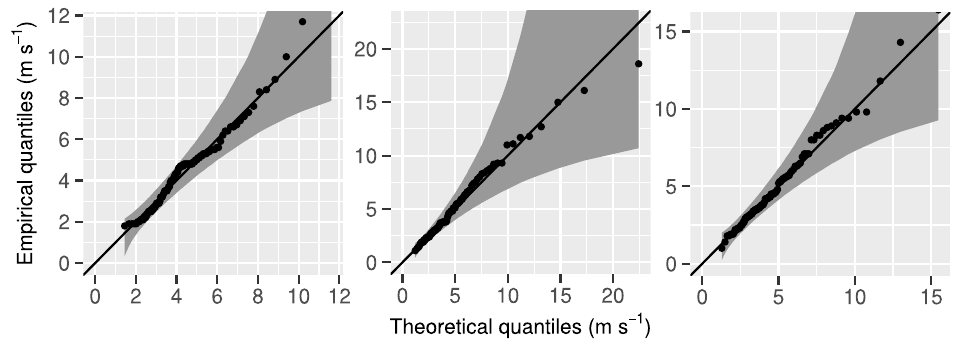}
    \caption{\centering Quantile-quantile plots of the GEV distributions fitted to the houly wind gusts observed from June 22, 2016 to July 13, 2016, at the SwissMetNet stations in Geneva (left), Lugano (middle), and Zurich (right).}
    \label{fig:qqplot}
\end{figure*}

Figure \ref{fig:qqplot} demonstrates that the GEV distribution is well-suited for modelling the time series of wind gusts at the three selected stations. This suitability extends more broadly across our entire dataset as, in most cases, a Kolmogorov-Smirnov test performed at the $5 \%$ level does not reject the assumption of a GEV distribution. The p-values are  0.16, 0.77, and 0.99 for Geneva, Lugano and Zurich, respectively.

\subsection*{Sequential Feature Selection of meteorological variables} \label{seqfeatsel}

Before training any neural networks, a sequential feature selection was performed on the training set to identify relevant meteorological variables. To do so, generalised linear models (see next subsection for more details) were trained on each upper variable for three pressure levels (1000 hPa, 850 hPa, and 500 hPa) and on each surface variable in a forward selection scheme. The wind magnitude was calculated from the zonal and meridional wind fields as the focus lies on the value of wind gusts, rather than their direction. To account for dataset inhomogeneity, five sequential feature selections were performed. The first four variables from the selections were then used as features for generalised additive models and Neural Networks.

\begin{figure}[h!]
    \centering
    \includegraphics[width=0.8\linewidth]{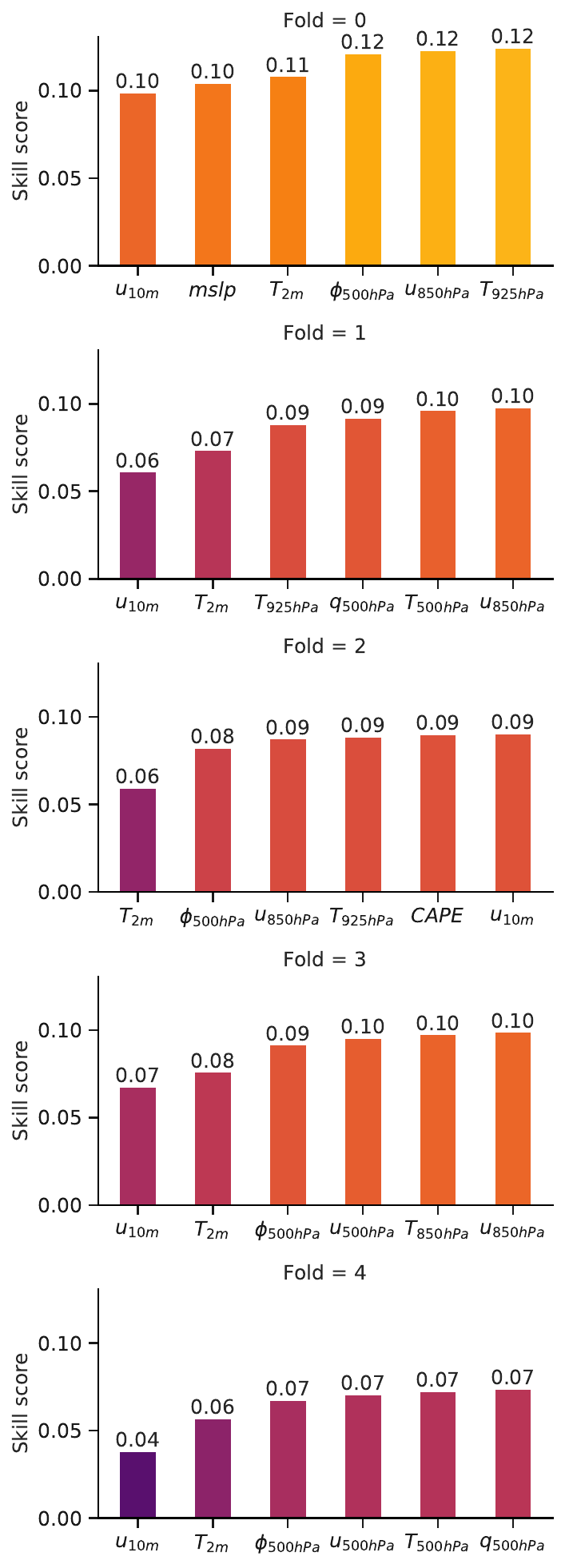}
    \caption{\centering Result of the sequential feature selection for each fold. The skill score refers to the maximum likelihood ratio between the VGLM result and the GEV-approximated climatology.}
    \label{fig:seqfeatsel}
\end{figure}

Figure \ref{fig:seqfeatsel} shows the results of the feature selections: the first four variables are wind speed at 10 m ($u_{10m}$), temperature at 2 m ($T_{2m}$), geopotential at 500 hPa ($\phi_{500hPa}$), and wind at 850 hPa ($u_{850hPa}$). It is unsurprising: wind at 10 m gives a mean value that wind gusts exceed, temperature and geopotential give information about small-scale and large-scale pressure systems, and wind at 850 hPa can be interpreted as the influence of lower-atmospheric winds on surface wind gusts. More surprising is that none of the storm-related variables that were computed from Pangu-Weather outputs such as CAPE, the Level of Free Convection (LFC), or SRH, were selected in the first four variables. This could be linked to these variables being good predictors of a storm presence, but not necessarily of the wind gusts a storm induces, which are more related to Downward Convective Available Potential Energy (DCAPE), storm motion and entrainment of dry air \cite{markowski2010hazards}.

\subsection*{Generalised Additive Models}

Following \cite{friederichs_forecast_2012}, we first train Vector Generalised Linear and Additive Models (VGLMs and VGAMs) to predict the parameters of wind gust GEV distributions from Pangu-Weather output.

VGLMs generalize linear regression by modelling a linear relationship between feature vectors $\mathbf{x}$ and targets $\mathbf{y}$. Optionally, a predefined link function $g$ (e.g., logarithm or exponential) can be applied pointwise to $\mathbf{y}$. This is useful, for instance, to enforce positivity constraints.

VGAMs extend VGLMs by allowing the linear terms to be replaced with flexible, non-linear functions (e.g., splines), accounting for non-linear relationships between $\mathbf{x}$ and $\mathbf{y}$.

For distributional regression of the GEV distribution, the model is specified as:

\begin{equation}
    \begin{split}
        \mu(\mathbf{x}) &= \beta_\mu + f_\mu^{(1)}(x^{(1)}) + \cdots + f_\mu^{(n)}(x^{(n)}), \\
        \ln \sigma(\mathbf{x}) &= \beta_\sigma + f_\sigma^{(1)}(x^{(1)}) + \cdots + f_\sigma^{(n)}(x^{(n)}), \\
        s(\xi) &= \beta_\xi.
    \end{split}
\end{equation}

Here, the logarithm ensures the scale parameter $\sigma$ remains positive. The shape parameter $\xi$ is kept constant—i.e., independent of the features $\mathbf{x}$—to prevent numerical instability \cite{coles_introduction_2001}. To further ensure stability, $\xi$ is constrained to the interval $(-0.5, 1)$ using the function $s$, defined as the inverse of the sigmoid function.

The functions $f_{\cdot}^{(j)}$, for $j = 1, \ldots, n$, are either linear mappings $x \mapsto \alpha_j x$ (in VGLMs) or splines with three degrees of freedom (in VGAMs).

Here, the VGLM and VGAM were trained following a k-fold cross-validation scheme, using years 2016 to 2020 as training folds, and keeping the year 2021 for evaluation. Years 2016 to 2020 were filtered to contain 50\% of dates with detected convective storms over Switzerland, and the year 2021 was filtered to contain 100\% of dates with the same criterion. Although our primary focus is on thunderstorm-induced wind gusts (hence the 100\% of dates in the test set), incorporating wind gusts unrelated to thunderstorms in the training dataset allows the model to learn additional generalizable features of wind gusts. The models were trained on the cluster of the University of Lausanne, with 64G of RAM and 16 CPUs, and it took 1h 15m for the VGLM and 3h 30m for the VGAM. 

Following \cite{friederichs_forecast_2012}, we first trained station-wise VGAMs and VGLMs. In total, 165 models were fitted: one for each of the 33 lead times and each of the 5 clusters. The models would thus predict different $\mu$ and $\sigma$ for each station but fixed $f_{.}^{(j)}$ for each cluster and lead time. The features were in this case the interpolation of Pangu-Weather outputs on each station.

Aiming to fit one distribution per cluster, the input covariates were replaced by the mean value of each feature on the cluster (average of the interpolated values at the stations). Interestingly, the cluster-wise VGAM outperformed the station-wise VGAM (CRPSS of 0.41 against 0.26). It could be explained by the limited spatial resolution of Pangu-Weather, whose global-scale design lacks the small-scale spatial accuracy needed to benefit fully from station-specific modeling, and the unpredictability of individual stations, making individual forecasts less precise - even when forecast is a distribution. Additionally, using cluster-averaged features allowed us to avoid the complexity and computational cost of fitting 5 VGAM with approximately 120 covariates (4 variables across about 30 stations for each cluster). In the results section, only the cluster-wise VGAM is presented as it gave the best results. The same reasons ultimately lead us to prefer cluster-wise fits rather than station-based ones for the NNs to tackle Pangu inherent coarse resolution (being a global model), individual station unpredictability, and computational training and running cost.

\begin{figure}[h!]
    \centering
    \includegraphics[width=\linewidth]{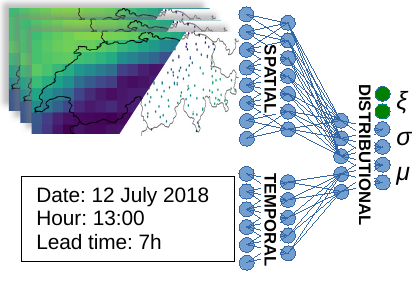}
    \caption{\centering NNs are composed of 3 blocks: a spatial block which takes as input either the full meteorological maps or their interpolation on the stations, a temporal block, and a distributional block which aggregates the results of the two preceding. }
    \label{fig:NNs}
\end{figure}

\begin{table*}[h!]
    \centering
    \begin{tabular}{|c|c|c|c|c|} \hline 
         Model Name&  VGAM&  ANN&  ViT& CNN\\ \hline 
         Type&  Generalized Linear Model&  Neural Network&  Neural Network& Neural Network\\ \hline 
         Inputs&  Station-based&  Station-based&  Gridded& Gridded\\ \hline 
         CRPS&  $1.785\pm0.03$&  $1.603\pm0.060$&  $1.605\pm0.056$& $1.597\pm0.053$\\ \hline
    \end{tabular}
    \caption{Summary - Comparison of the four post-processing models}
    \label{tab:comp}
\end{table*}

\subsection*{Convolutional neural network and vision transformer}

The cluster-wise VGAM does not take full advantage of the spatial information provided by Pangu-Weather: it only uses the average of the four previously defined variables on each cluster. However, the spatial structure of the different fields could also contain important information to predict wind gusts. Feeding pixel-wise features from the full Swiss domain into a VGAM is computationally prohibitive, as a $20 \times 34$ grid with 4 features per pixel results in $2720$ covariates. GPU-computation provides a better way to process large amounts of data in parallel through NNs.

Three different NNs were used with a similar global structure represented in Figure \ref{fig:NNs}:
\begin{itemize}
    \item The \textbf{spatial block} processes either Pangu-Weather outputs clipped to Switzerland (for CNN and ViT models) or interpolated values at the 128 SwissMetNet stations. Its structure depends on the specific architecture.
    \item The \textbf{temporal block} encodes time using four variables: normalised day of year, cosine and sine of normalised hour of day, and normalised lead time. It consists of a single-layer perceptron with 64 neurons and Leaky ReLU activation.
    \item The \textbf{distributional block} aggregates the spatial and temporal blocks' output to predict the GEV parameters $\mu$, $\sigma$, and $\xi$ for each cluster. It contains two fully connected layers (64 and 10 neurons). A SoftPlus activation ensures $\sigma > 0$, while the shape parameters $\xi_1, \ldots, \xi_5$ are learned but not input-dependent.
\end{itemize}

Table~\ref{tab:comp} summaries the key aspects of each of the four post-processing models, and table~\ref{fig:model_structure} details the NN architectures. For the Vision Transformer, only the hyperparameters are shown as we strictly follow the implementation in \cite{noauthor_tutorial_nodate}. As previously, the 2016–2020 data were filtered to retain 50\% of days with detected convective storms over Switzerland. In contrast, 2021 includes all such days (100\%) using the same detection criterion.

\begin{table}[h]
\centering

\begin{tabular}{|c|c|c|}
\hline
\multicolumn{3}{|c|}{\textbf{CNN}} \\
\multicolumn{3}{|c|}{(Total: 2,352 -- 9,408 bytes)} \\
\hline
\textbf{Name} & \textbf{Shape} & \textbf{Size} \\
\hline
Conv\_0/bias  & (16,)         & 16   \\
Conv\_0/kernel& (2, 2, 4, 16) & 256  \\
Conv\_1/bias  & (32,)         & 32   \\
Conv\_1/kernel& (2, 2, 16, 32)& 2,048 \\
\hline
\end{tabular}

\vspace{1cm} 

\begin{tabular}{|c|c|c|}
\hline
\multicolumn{3}{|c|}{\textbf{ANN}} \\
\multicolumn{3}{|c|}{(Total: 38,304 -- 153,216 bytes)} \\
\hline
\textbf{Name} & \textbf{Shape} & \textbf{Size} \\
\hline
Dense\_0/bias  & (256,)    & 256   \\
Dense\_0/kernel& (4, 256)  & 1,024 \\
Dense\_1/bias  & (128,)    & 128   \\
Dense\_1/kernel& (256, 128)& 32,768 \\
Dense\_2/bias  & (32,)     & 32    \\
Dense\_2/kernel& (128, 32) & 4,096 \\
\hline
\end{tabular}

\vspace{1cm} 

\begin{tabular}{|c|c|}
\hline
\multicolumn{2}{|c|}{\textbf{Vision Transformer}} \\
\multicolumn{2}{|c|}{(Total: 412,160 -- 1,648,640 bytes)} \\
\hline
\textbf{Parameter}   & \textbf{Value} \\
\hline
embed\_dim           & 128           \\
hidden\_dim          & 256           \\
num\_heads           & 8             \\
num\_channels        & 3             \\
num\_layers          & 3             \\
patch\_size          & 4             \\
num\_patches         & 50            \\
dropout\_prob        & 0.2           \\
\hline
\end{tabular}

\caption{Model Structure: CNN, ANN, and Vision Transformer}
\label{fig:model_structure}
\end{table}

\subsection*{Training}

All NN models were trained using ADAM optimizer. A grid search was performed to tune their hyperparameters (learning rate, number of layers/attention layers, number of neurons/attention head on each layer). Additionally, a Bayesian hyperparameter research was performed on the CNN, but it was not able to beat the grid search-obtained CNN. Training was always performed on a maximum of 50 epochs with early stopping, and the final parameters were computed as the mean of the three best states from training.

In the case of the ANN and the CNN, adding L2 regularisation to CRPS loss with a coefficient $\alpha = 10$ performed better than no regularisation or L1 regularisation.

\subsection*{Interpretation of CRPS performance v. CRPSS performance}

CRPS and CRPSS should not be interpreted in the same way, and a better CRPS does not necessarily imply a better CRPSS. While CRPS typically reflects the absolute predictability of one model or region compared to another, CRPSS highlights the relative improvement in predictability of a model on a specific test set. By normalizing against a reference forecast (typically climatology), CRPSS provides a more balanced assessment of model performance, reflecting improvements relative to local predictability.

Using CRPS to compare model performance across regions with different meteorological characteristics can lead to misleading conclusions: a worse score in one region may result from its inherent unpredictability. A good model helps mitigate this unpredictability by improving forecast skill. Even if the model still underperforms compared to other regions in absolute terms, the improvement can be more significant in already unpredictable areas.

\begin{figure}[h!]
    \centering
    \includegraphics[width=\linewidth]{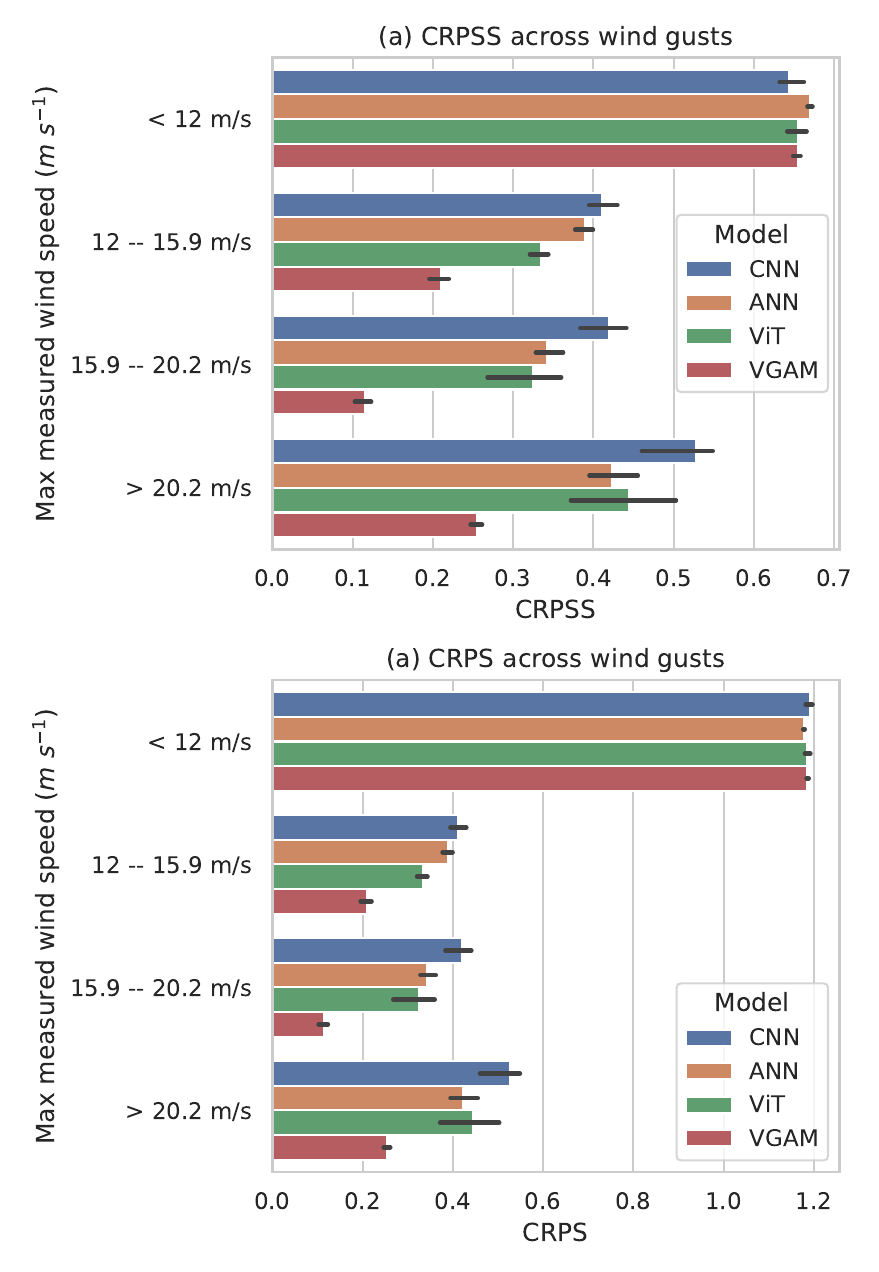}
    \caption{\centering Better CRPSS performance is not always related to better CRPS performance.  Panel (a) is taken from Figure~\ref{fig:quantiles}, while panel (b) provides the same comparison in terms of CRPS. On panel (a), best models have a high CRPSS. On panel (b), best models have a low CRPS. An inversion is observed in model ranking.}
    \label{fig:crpsVcrpss}
\end{figure}

To illustrate these differing interpretations, consider the case of filtered maximum observed gust speed. Figure \ref{fig:crpsVcrpss} shows an inversion in performance rankings depending on whether CRPS or CRPSS is used. This is expected: when filtering by observed gust strength, biases can emerge due to regional differences in average wind conditions. Such filtering may inadvertently compare model performance across regions with inherently different climatological characteristics, again leading to potentially misleading conclusions. CRPSS helps mitigate this bias by incorporating regional climatology into the evaluation: in regions with stronger average gusts -- where wind gust speed is more dispersed hence less predictable -- our post-processing models achieve greater relative improvement in forecast performance than in regions where wind gust predictability is already high becasue of lower average wind speed and lower dispersion (see Supplementary Figures 6 and 7, SI section S4 for more details about regional difference in CRPS and CRPSS).

\section*{Data availability}

The weights and biases of the trained data-driven models are available at \url{https://doi.org/10.5281/zenodo.15092539}. The Pangu-Weather model was run using ECMWF’s AI models implementation at \url{https://github.com/ecmwf-lab/ai-models}. Storm tracks were derived from a publicly available radar-based product \url{https://zenodo.org/records/6534510}. SwissMetNet and ICON-CH2-EPS data can be obtained free of charge for academic research by contacting MeteoSwiss customer service \url{https://www.meteoswiss.admin.ch/about-us/contact/contact-form.html}.

\section*{Code availability}
The underlying code for this study is available in \url{https://github.com/AntLrc/ConvectiveThunderstormWindGust_ML} and can be accessed via this persistent link \url{https://github.com/AntLrc/ConvectiveThunderstormWindGust_ML/releases/tag/v1.1}.

\section*{Acknowledgments}

EK would like to thank the Expertise Center for Climate Extremes (ECCE) at the University of Lausanne for financial support. TB acknowledges partial support from the Swiss National Science Foundation (SNSF) under Grant No. 10001754 (``RobustSR'' project). The funder played no role in study design, data collection, analysis and interpretation of data, or the writing of this manuscript. MeteoSwiss provided the SwissMetNet and ICON-CH2-EPS data. We thank Valerie Chavez, Milton Gomez, Louis Poulain-Auzeau, Margot Sirdey, Flavio Calvo, and Mauricio Lima for their helpful advice, which improved the quality of the post-processing models.

\section*{Author contributions}

AL, EK, MF, and TB contributed to the conceptualization of the study. Data curation was carried out by AL, DN, and MF. Formal analysis was conducted by EK, TB, and AL. AL developed the software and prepared the original draft of the manuscript. All authors contributed to the investigation, methodology, validation, visualization, and to the review and editing of the manuscript. TB and EK provided project supervision and administration. TB secured the necessary resources and funding for the project. All authors have read and approved the manuscript.

\section*{Competing interests}

The authors declare no competing interests.

\pagebreak
\printbibliography[heading=bibintoc]

\end{document}


\renewcommand{\figurename}{Supplementary Figure}

\title{Supplementary information \\ \Large Improving Predictions of Convective Storm Wind Gusts through Statistical Post-Processing of Neural Weather Models}
\date{}
\maketitle

\section*{S1: Oscillations in model performance}

\begin{figure}[h]
    \centering
    \includegraphics[width=0.8\linewidth]{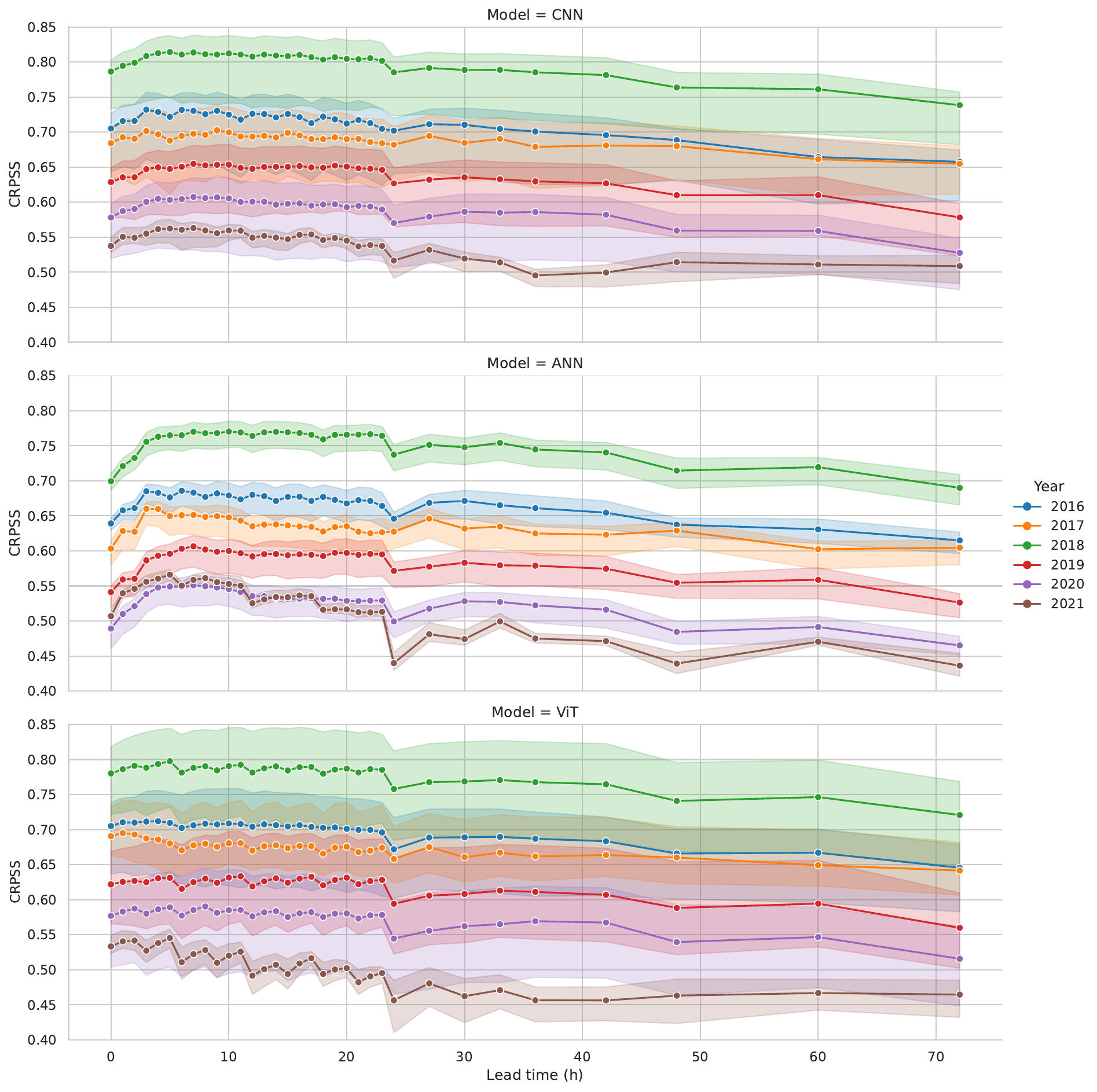}
    \caption{CRPSS by lead time for the CNN, ANN, and ViT, shown separately for 2016–2021. Two robust features appear across all years and methods: (i) a sawtooth pattern for lead times $<24$~h, caused by different short-range step compositions in Pangu-Weather; and (ii) a drop at 24~h, caused by the switch to the 24~h submodel.}
    \label{fig:oscil}
\end{figure}

At first glance, the CRPSS in Fig.~2b oscillates as a function of lead time. Here we clarify what we mean by ``oscillations'', examine them year by year, and explain why they arise from the Pangu-Weather architecture and, to a lesser extent, from choices in the post-processing models. Although sampling effects might be suspected, Supplementary Fig.~\ref{fig:oscil} shows that the same patterns recur in every year, which argues against sampling as the primary cause. 

Two distinct features are present:
(i) a sawtooth behaviour for lead times below 24~h, and
(ii) a discrete drop in skill at exactly 24~h. Both appear across years (2016–2021) and across post-processing methods (CNN, ANN, ViT).

First, Pangu-Weather uses different submodels at different horizons and composes them to reach a target lead time with a minimal number of steps \cite{bi_pangu-weather_2022}, with oscillations also reported by \cite{zhong2024fuxi}. Adjacent lead times can therefore be produced by different step compositions (e.g., $6{+}1$~h vs.\ $3{+}3{+}1$~h), which changes error characteristics slightly from one lead time to the next. This architectural choice explains the sawtooth pattern below 24~h.

Second, at 24~h the system switches to the dedicated 24~h submodel. That switch introduces a change in error properties and coincides with a notable decrease in skill that is visible in all years and methods. Because these two features are aligned with the Pangu submodel boundaries and persist across years with different storm samples, they are best attributed to the driving neural weather model rather than to sampling variability.
 
Post-processing models largely transmit these Pangu biases but can modulate their amplitude. In the CNN and ANN, L2 regularisation and an explicit temporal block keep temporal encodings in our post-processing models influential, which tends to smooth the sawtooth. In contrast, the ViT (dropout in the spatial block, no L2) relies more heavily on spatial features and mirrors Pangu’s lead-time structure more closely. The VGAM, trained independently at each lead time, reproduces the Pangu/ViT pattern most directly because it does not share information across lead times. 

In summary, the pre-24~h sawtooth arises from different short-range step compositions in Pangu-Weather, while the 24~h jump reflects the switch to the 24~h submodel. Post-processing does not create these patterns; it mainly inherits them, with minor smoothing when temporal conditioning and regularisation are stronger.

\section*{S2: CRPSS stabilisation for longer lead times}

Figure~2b also shows a stabilisation of the CRPSS at longer lead times, while the CRPS increases with lead time. This behaviour arises because CRPSS is a normalised skill score relative to climatology: when both the forecast CRPS and the climatological CRPS degrade at similar rates, the relative improvement can plateau even as absolute errors grow. In our case, the plateau reflects persistent medium-range skill in the Pangu-Weather driver, particularly for 10~m wind fields that strongly inform our wind gust predictions.

\begin{figure}[h]
    \centering
    \includegraphics[width=1\linewidth]{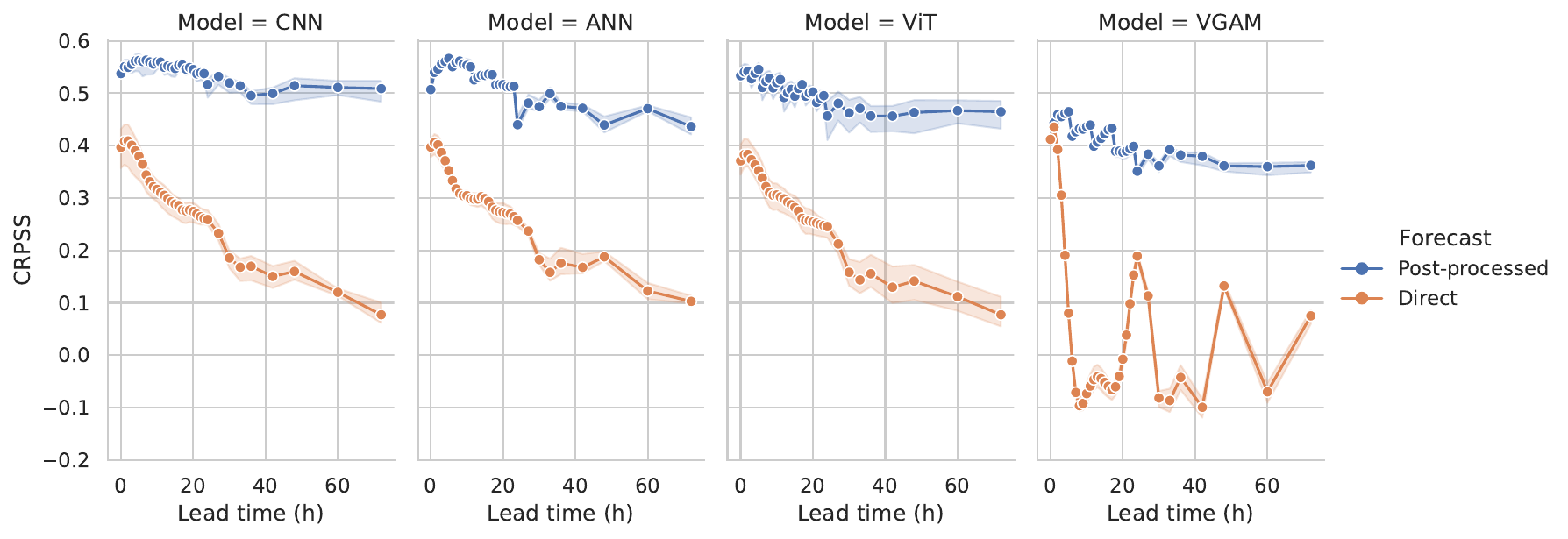}
    \caption{\centering CRPSS by lead time for each model (VGAM, ANN, ViT, CNN): direct forecasts from ERA5-derived initial conditions versus post-processing of Pangu-Weather. For VGAM, the curve dips below zero at some lead times. Post-processed Pangu-Weather retains positive skill and tends to plateau by 48–72~h, whereas direct baselines approach CRPSS~$=0$.} 
    \label{fig:comparison-direct}
\end{figure}

This interpretation is supported by a comparison to direct forecasts based only on ERA5-derived initial conditions. If the CRPSS plateau were independent of the driver, similar stabilisation would appear for the direct baselines. Instead, as Supplementary Figure~\ref{fig:comparison-direct} shows (and as in Figure~2c of the main text), at long lead times the CRPSS of direct forecasts that do not post-process Pangu-Weather tends towards zero. By contrast, Pangu-based post-processing maintains positive CRPSS through 48–72~h.

Due to computational constraints, we did not train beyond 72~h. At longer horizons we expect CRPSS to decline as the driver’s predictability decreases, consistent with the trend observed for our direct forecasting baselines.

\section*{S3: Performance for severe gusts: Sensitivity to filtering procedure}

\begin{figure}[h!]
    \centering
    \includegraphics[width=.5\linewidth]{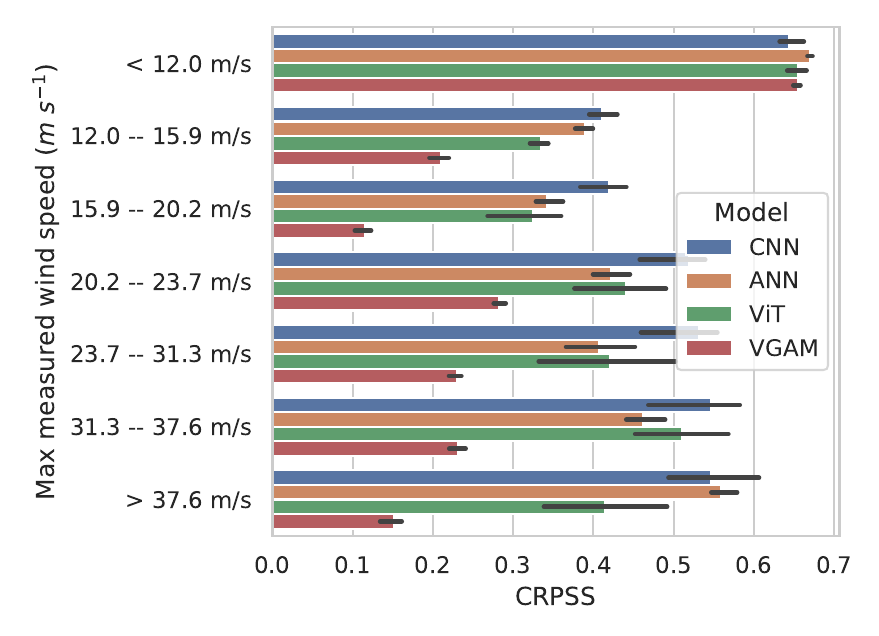}
    \caption{\centering CRPSS for filtered wind gusts based on the maximum wind gust observed in the cluster at each time step.}
    \label{fig:max-gust}
\end{figure}

\begin{figure}[h!]
    \centering
    \includegraphics[width=.5\linewidth]{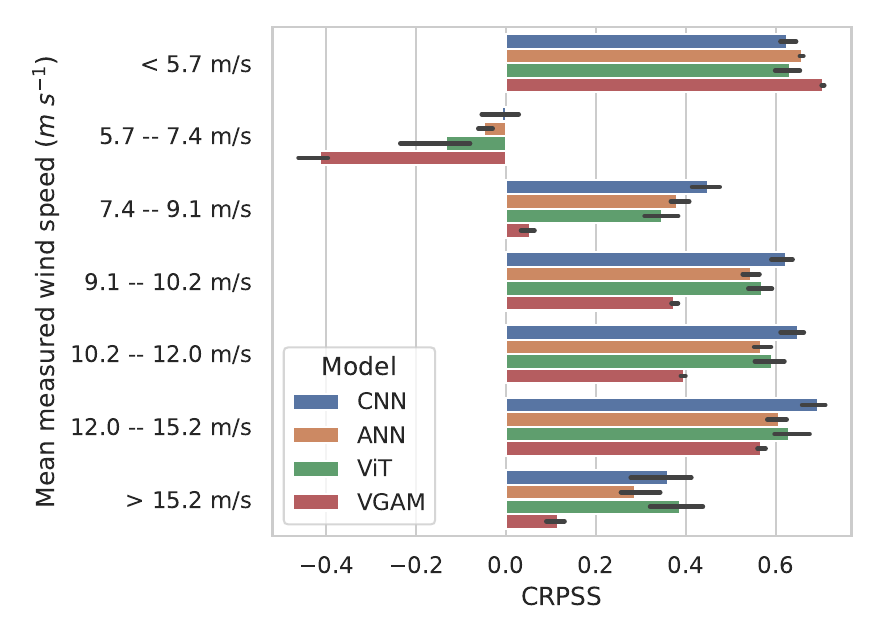}
    \caption{\centering CRPSS for filtered wind gusts based on the mean wind gust observed in the cluster at each time step.}
    \label{fig:mean-gust}
\end{figure}

\begin{figure}[h!]
    \centering
    \includegraphics[width=.5\linewidth]{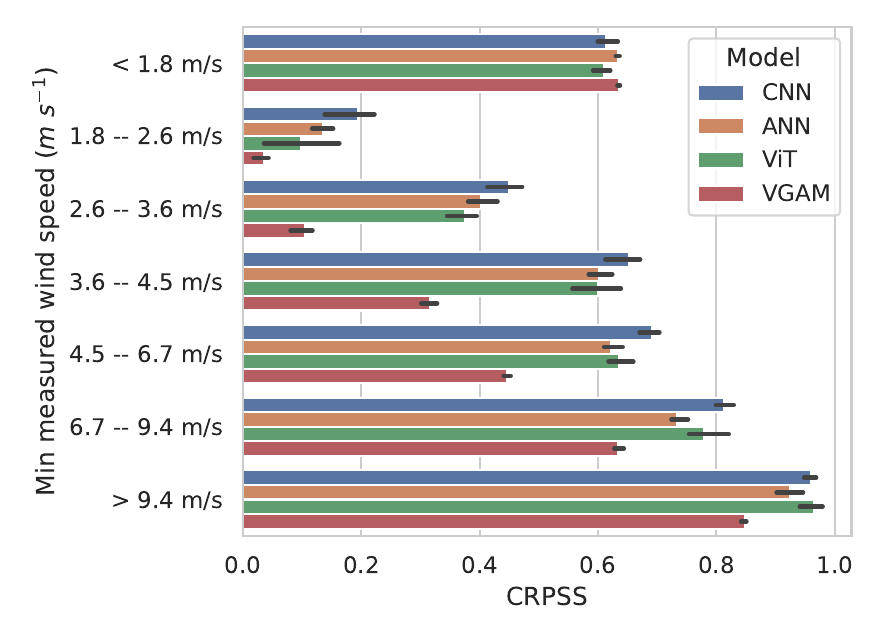}
    \caption{\centering CRPSS for filtered wind gusts based on the minimum wind gust observed in the cluster at each time step.}
    \label{fig:min-gust}
\end{figure}

Post-processing model performance across wind gust speeds was computed for different filters. Supplementary Figures \ref{fig:max-gust}, \ref{fig:mean-gust} and \ref{fig:min-gust} illustrate that most models yield positive CRPSS across gust values, which means they beat climatology even in extreme situations. Thresholds are always based on the following quantile levels: $0.5$, $0.75$, $0.9$, $0.95$, $0.99$, $0.999$.

\section*{S4: Performance across regional clusters}

\begin{figure}[h!]
    \centering
    \includegraphics[width=1\linewidth]{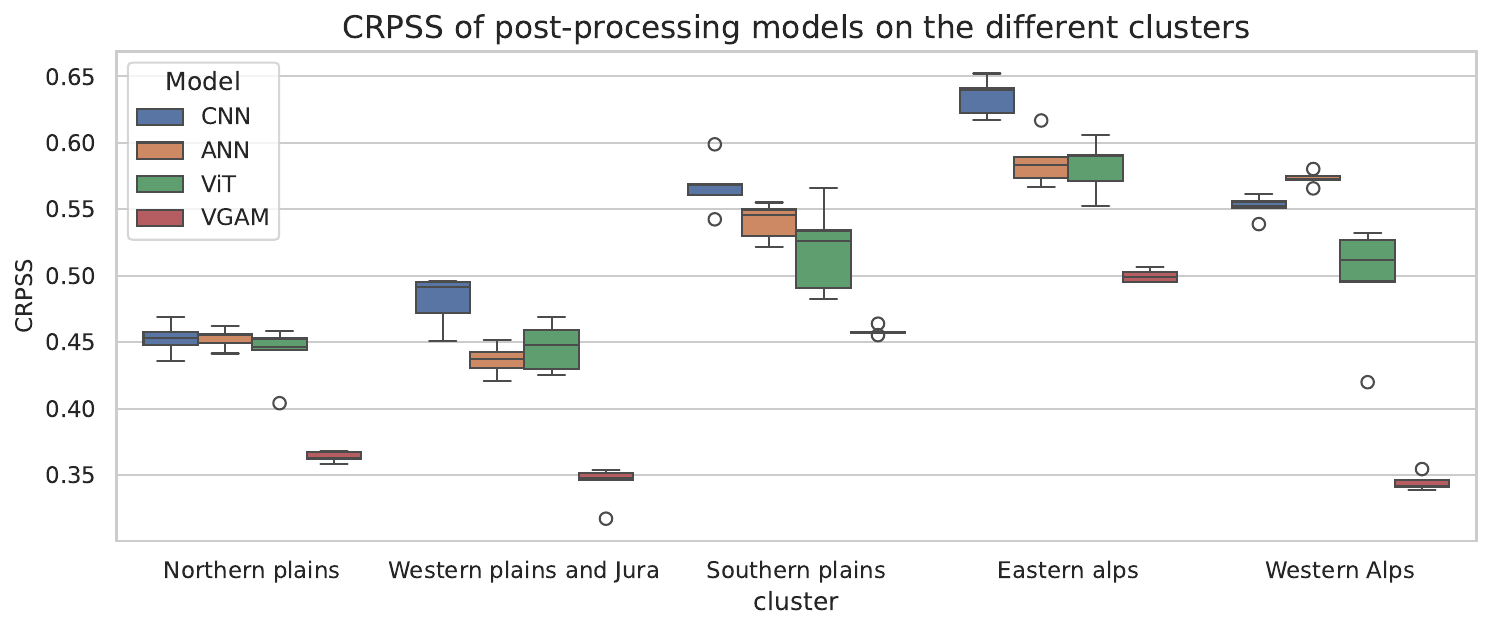}
    \caption{\centering CRPSS by region. Post-processing improves upon local climatology more over the Alps and southern plains.}
    \label{fig:cluster-crpss}
\end{figure}

\begin{figure}[h!]
    \centering
    \includegraphics[width=1\linewidth]{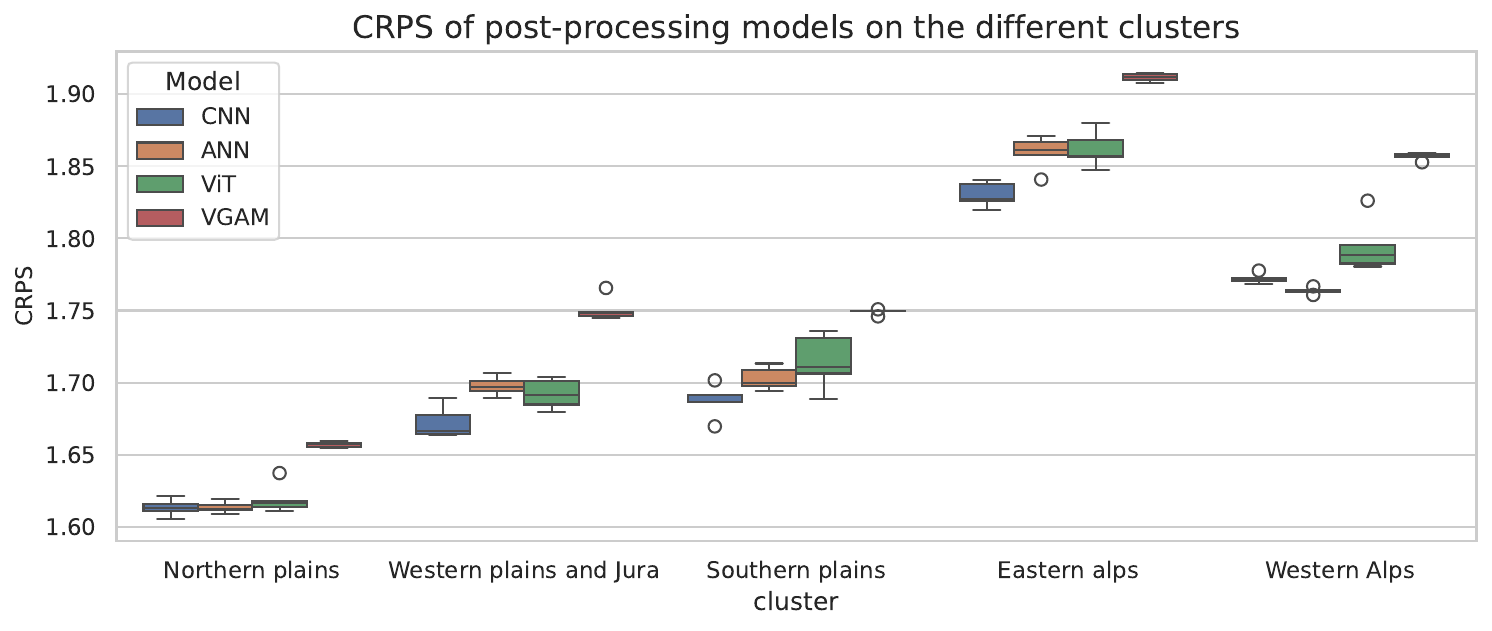}
    \caption{\centering CRPS by region (lower is better). Errors are larger over the Alps and southern plains.}
    \label{fig:cluster-crps}
\end{figure}

Relative to local climatology, Supplementary Figure~\ref{fig:cluster-crpss} shows that the post-processing models (except VGAM) achieve higher CRPSS over the Alps and southern plains than over the northern and western plains. This suggests that climatology is more dispersed over the Alps and southern plains—where extremes are more frequent—than north of the Alps. However, Supplementary Figure~\ref{fig:cluster-crps} shows that in absolute terms (CRPS; lower is better), models perform better in northern Switzerland, where conditions are more predictable.

\section*{S5: Discussion: Choice of the test year}

Although only one test year might raise questions regarding the limited size of the test set, several reasons led to this choice that we believe is reasonable.

\begin{figure*}[h!]
    \centering
    \includegraphics[width=1\linewidth]{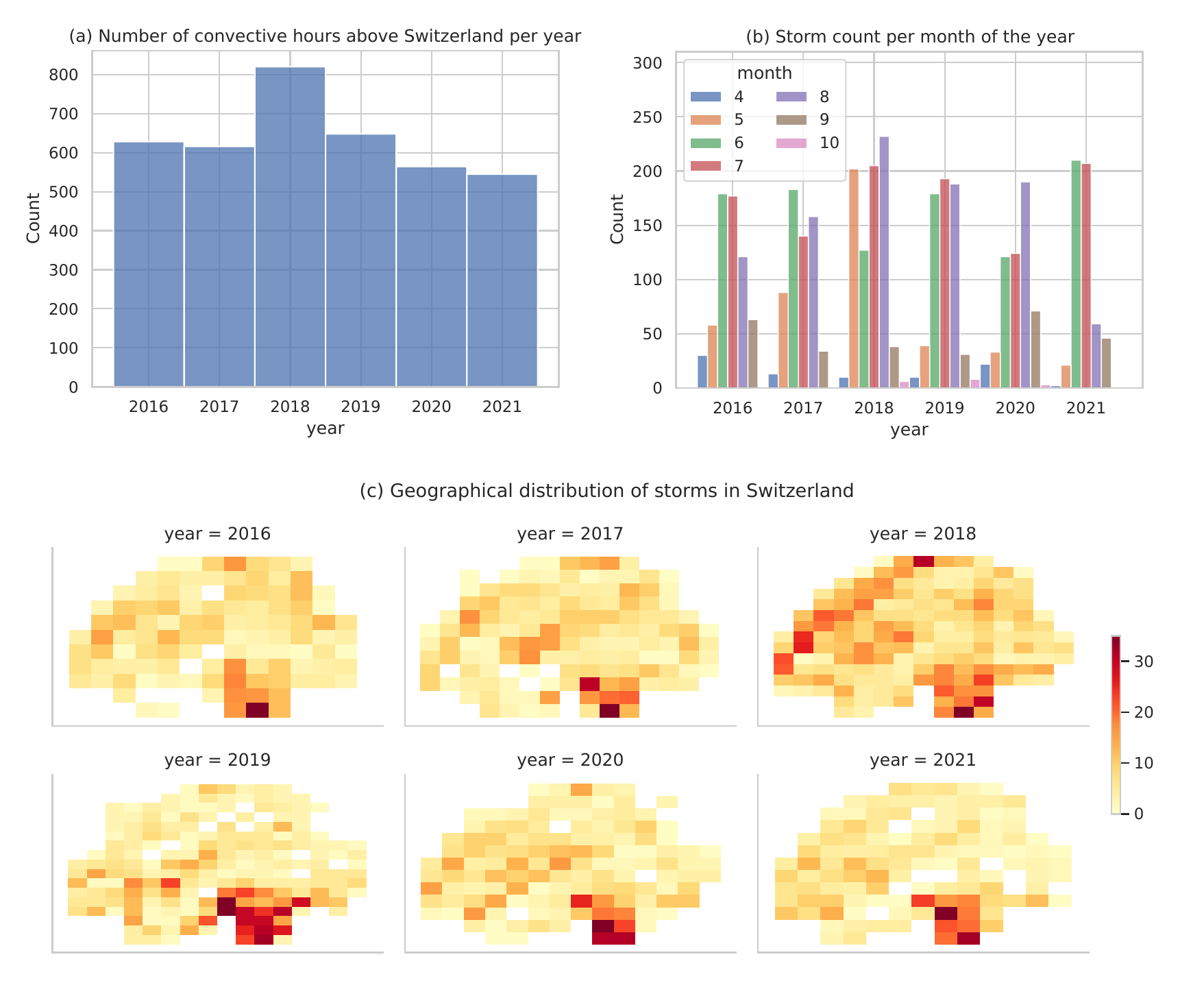}
    \caption{2021 is comparable to years 2016--2020 in terms of number of hours (a), distribution by month (b), and location (c) of convective events.}
    \label{fig:dist-storms}
\end{figure*}

First, Figure~\ref{fig:dist-storms} illustrates that the distribution of convective storm days in 2021 is comparable to those in the 2016--2020 training period, indicating that it is climatologically representative in terms of storm frequency.

Second, while 2021 was average in frequency, it nonetheless included several of the most damaging convective events in recent decades, such as the June 28 and July 8 hailstorms \cite{storms2021june}. These high-impact events stress test the model's performance.
    
Third, to mitigate sampling variability, we trained all models using 5-fold cross-validation over the 2016--2020 period. Figures~2 and~3 already report the mean and min/max skill across folds.
    
Finally, ICON reforecasts and complete SwissMetNet observations were only available for 2021, which constrained our choice of test year for consistency in our benchmarking across models.

\printbibliography[heading=bibintoc]